# Picosecond acoustic excitation driven ultrafast magnetization dynamics in dielectric Bi-substituted yttrium iron garnet

Marwan Deb[1*], Elena Popova[2], Michel Hehn[1], Niels Keller[2], Stéphane Mangin[1], Gregory Malinowski[1]

[1]Institut Jean Lamour (IJL), CNRS UMR 7198, Université de Lorraine, 54506 Vandœuvre-lès-Nancy, France.

[2]Groupe d'Etude de la Matière Condensée (GEMaC), CNRS UMR 8635, Université de Paris-Saclay, 78035 Versailles, France.

**Abstract:**

Using femtosecond optical pulses, we have investigated the ultrafast magnetization dynamics induced in a dielectric film of bismuth-substituted yttrium iron garnet (Bi-YIG) buried below a thick Cu/Pt metallic bilayer. We show that exciting the sample from Pt surface launches an acoustic strain pulse propagating into the garnet film. We discovered that this strain pulse induces a coherent magnetization precession in the Bi-YIG at the frequency of the ferromagnetic resonance. The observed phenomena can be explain by strain-induced changes of magnetocrystalline anisotropy via the inverse magnetostriction effect. These findings open new perspectives toward the control of the magnetization in magnetic garnets embedded in complex heterostructure devices.

*Corresponding author: marwan.deb@univ-lorraine.fr

**I. Introduction:**

The experimental discovery of the subpicosecond demagnetization of Ni films following the excitation by 60 femtosecond optical pulse [1] has opened a new and rapidly growing research field of modern magnetism called femtomagnetism [2,3]. An ultimate goal of this field is to control the magnetization at the fastest possible speed and in the most efficient way. To this end, intense researches are being carried out to investigate laser-induced ultrafast magnetic process and understanding the fundamental mechanism behind their excitation. It was demonstrated that ultrashort optical pulses can trigger in magnetic materials various important magnetic processes, including magnetic phase transition [4-6], magnetization switching [7,8], as well as a coherent spin precession [9,10]. In metallic materials, thermal effects resulting from the energy absorbed by the medium play a crucial role in laser-induced magnetic phenomena [11]. On the other hand, the possibility of controlling the spin precession in dielectric with light was demonstrated via non-thermal mechanisms like the inverse Faraday effect [10,12].

Recently, the field of femtomagnetism started investigating alternative ways to control the magnetization using other ultrashort stimuli including hot-electron pulses [13-15], Terahertz pulses [16,17], as well as acoustic pulses [18-20]. This is due to two main reasons. The first is to improve the fundamental understanding in highly debated issues related to ultrafast magnetic phenomena induced by optical pulses like the ultrafast demagnetization [13,21] and reversal [14,15]. The second is to discover a versatile tool that allows non-thermal and ultrafast control and reversal of the magnetization in metal, semiconductor, and dielectric films and heterostructures. In particular, for the latter reason, the use of acoustic pulse can offers at least two advantages. Firstly, acoustic pulses can produce a high mechanical stress of several GPa [22]. As a result, a large modification of lattice parameter occurs and the magnetization can be therefore non-thermally changed via the inverse magnetostriction effect [18,19]. Secondly, acoustic pulses have a large propagation distance of several millimeters with low energy dissipation [23,24]. Consequently, they provide opportunity for non-thermal manipulation of spins in films deeply embedded in opaque hetorestructure devices. In 2010, the control of the magnetization dynamics by a picosecond acoustic pulse was demonstrated by Scherbakov et al. in GaMnAS ferromagnetic semiconductor [18]. Later, acoustically-induced magnetization dynamic was extended to ferromagnetic metals [19,25,26]. An important question in this context is the possibility to take advantage of picosecond acoustic pulse to trigger a magnetization dynamics in magnetic dielectrics.

In this paper we present the results of an experimental study exploring the laser induced ultrafast magnetization dynamics in bismuth-substituted yttrium iron (Bi-YIG) garnets buried below a thick Cu/Pt metallic bilayer. By exciting the Pt/Cu/Bi-YIG trilayers from Pt surface, we find out that an acoustic strain pulse is generated and propagated into the garnet film. We demonstrate that the strain pulse induces a coherent magnetization precession at the frequency of ferromagnetic resonance. The obtained results are explained by strain-induced change of magnetocrystalline anisotropy via the inverse magnetostriction effect. In addition, we demonstrate that we can control the magnetization precession amplitude by tuning the amplitude of the acoustic pulse.

The paper is organized as follows. First, we describe in Sec. II the experimental methods and the static magnetic and magneto-optical properties of the sample. Then, we present and discuss in Sec. III the experimental results of the time-resolved magneto-optical and reflectivity measurements as a function of the external magnetic field and the laser energy density. Finally, we summarize in Sec. IV our findings.

## II. SAMPLE PROPERTIES AND EXPERIMENTAL METHODS:

Bismuth-substituted yttrium iron garnet ($Bi_xY_{1-x}Fe_5O_{12}$, Bi-YIG) materials have the cubic Ia-3d space group, which is characterized by three different crystallographic sites (octahedral 16a, tetrahedral 24d, and dodecahedral 24c) formed by the oxygen atoms [27]. The non-magnetic Bi and Y atoms occupy the dodecahedral 24c sites, while the magnetic Fe atoms are distributed between the octahedral 16a and tetrahedral 24d sites. This two Fe sublattices are nonequivalent and coupled by a strong antiferromagnetic superexchange interaction, leading to a ferrimagnetic state with high Curie temperature ($T_C$ > 550 K). These materials have attracted a great deal of attention due to their fascinating proprieties at room temperature, such as the good transparency in the infrared and visible spectrum (gap ~ 2.5 eV) and the very large magneto-optical (MO) Faraday effect (~ $10^4$ deg/cm at 2.4 eV) [28], which make them suitable for MO recoding media and non-reciprocal MO devices [27,29]. Beside the technological importance of the large MO effects, they have also been used as an efficient tool to study fundamental science related to magnetism in Bi-YIG such as the spin dependent band structure [30] as well as light-induced ultrafast magnetization dynamics and switching [31-36].

The experiments were performed on 140-nm-thick film of $Bi_2Y_1Fe_5O_{12}$, grown by pulsed laser deposition onto a gadolinium gallium garnet ($Gd_3Ga_5O_{12}$, GGG) (100) substrate. The structural properties were characterized in-situ by reflection high-energy electron diffraction (RHEED) and ellipsometry, and ex-situ by X-ray diffraction and transmission electronic microscopy. The film is single phase and epitaxial with atomically sharp interface. The magnetic and magneto-optical (MO) properties of the film were investigated using a custom designed broad band MO spectrometer based on 90°-polarization modulation technique. Details on the experimental setup are described in Ref [32,37]. Figures 1(a) and 1(b) show the spectral dependency of the rotation ($\Theta_F$, $\Theta_K$) and ellipticity ($\varepsilon_F$, $\varepsilon_K$) Faraday and Kerr spectra measured at 300 K in polar configuration with saturating external magnetic field applied perpendicular to the film plane. $\Theta_F$ is negative above 487 nm with a minimum at 520 nm and positive between 487 nm and 353 nm with a maximum at 390 nm, whereas $\varepsilon_F$ shows two pics centered at 474 nm and 357 nm. For the MO Kerr signals, the largest absolute amplitudes occur in the vicinity of the optical band gap: $\Theta_K$ reaches -1.6 deg near 510 nm and $\varepsilon_K$ reaches -1.3 deg near 540 nm. We note that the Faraday and Kerr spectra show a good agreement with previous study of MO properties in Bi-YIG [28,37,38]. From a fundamental point of view, they are well described by the crystal field energy levels of $Fe^{3+}$ ions in tetrahedral and octahedral symmetries, which acquire a large enhancement in the spin-orbit splitting due to their hybridization with Bi-6s orbital [37,39,40]. This phenomenon is proposed to be at the origin of the increase of MO proprieties in Bi-YIG with increasing Bi content. Let us also mention that the positions of the peaks of $\Theta_F$ nicely fit into recent results showing the evolution of the energy level transition associated with the tetrahedral and octahedral iron sites as a function of Bi content in $Bi_xY_{3-x}Fe_5O_{12}$ ( $0.5 \leq x_{Bi} \leq 3$ ) [37], which confirm the bismuth concentration in our samples. The normalized polar and longitudinal Kerr hysteresis loops of the garnet film are shown in Figs. 1(c) and 1(d) respectively. The normalized remanence ($M_r/M_s$) is of 0.05 and 0.45 for the polar and longitudinal configuration respectively. The very weak polar remanence show that the easy axis of the magnetization is in the film plane.

In order to explore the effect of an ultrashort strain pulse on the magnetization dynamics in iron garnet, a thick Cu(100)/Pt(5) nonmagnetic metallic bilayer was deposited by dc magnetron sputtering on top of the garnet film (see Fig. 1(e)). The numbers in parentheses are in nanometers and represent the thickness d of the layer. Let us mention that each metallic layer plays a crucial role in our artificial structure. The Pt top layer is important due to its high electron-phonon coupling constant (109 $10^{16}$ $W.m^{-3}.K^{-1}$) [41] and absorption at 800 nm [13], which enables the generation of coherent acoustic phonon with quite high frequency [42,43].

The Cu layer is important due to its high hot-electron life time. Indeed, a thick Cu layer allows protecting the magnetic Bi-YIG film from the direct laser excitation, while the hot-electron can travel balistically for Cu thickness up to few hundreds of nanometers [13,14]. The arrival of the hot-electron at the back side of Cu modifies its reflectivity and can be therefore used as a mark of the zero time delay that defines the onset of the pump excitation [44].

The time-resolved MO and reflectivity measurements were performed at 300 K with the all-optical pump-probe configuration sketched in Fig. 1(c). Briefly, we have employed a femtosecond laser pulse issued from an amplified Ti-Sapphire laser system operating at a 5 kHz repetition rate and delivering 35 fs pulses at 800 nm to generate the pump and the probe beams. The pump beam is kept at the fundamental of the amplifier at 800 nm and excites the sample at normal incidence from the Pt side, while the probe beam is frequency doubled to 400 nm using a barium boron oxide crystal and incident with a small angle of 6° onto the GGG substrate. Both beams are linearly polarized and focused onto the sample in spot diameters of ~260 µm for the pump and ~60 µm for the probe. The probe wavelength is well below the optical absorption edge of the GGG [45], which allows the probe to penetrate the substrate and reach the Bi-YIG layer. After interaction with the Bi-YIG, the reflected probe pulses allow measuring the differential changes of the MO polar Kerr rotation $\Delta\Theta_K$ (t) and reflectivity $\Delta R$ (t) induced by the acoustic pulse as a function of the time delay t between the pump and probe pulses using a synchronization detection scheme. The external magnetic field $H_{ext}$ is applied perpendicular to the plane of the film.

### III. RESULTS AND DISCCUSSION:

Figure 2(a) shows the time resolved MO Kerr effect (TR-MOKE) measurement of the dynamics induced by a laser energy density of 11.3 mJ cm$^{-2}$ for $H_{ext}$ = 3.3 kG. We note that the TR-MOKE signal changes its sign when the direction of $H_{ext}$ is reversed. In addition, the zero time delay corresponds to the arrivals of hot-electron pulse to the back side of Cu layer, as revealed by the TR-MOKE signals measured in areas with and without the Pt/Cu bilayers. On the other hand, a strong peak in the TR-MOKE signal appears at t = 40 ps, which is very close to the time t = $d_{Cu}/V_{Cu} + d_{Bi-YIG}/V_{Bi-YIG} \approx 41$ ps required for an acoustic pulse to cross the Cu and Bi-YIG layers, where d and V are the thickness and longitudinal sound velocity characterizing Cu and Bi-YIG and their values are $d_{Cu}$ = 100 nm, $d_{Bi-YIG}$ = 140 nm, $V_{Cu}$ = 4730 m/s [13] and $V_{Bi-YIG} \approx 6700$ m/s [46]. The changes observed in $\Delta\Theta_K$ (t) signal in the time delay between 0 and 40 ps have a nonmagnetic origin. This phenomenon is the same as reported in Ref [47] for semiconductor as it shows the same characteristic behaviors: (i) The variation of its amplitude with the magnetic field is the same as the static MO response of the sample, i.e, it saturates for $H_{ext}$ higher than the saturating field $H_{sat}$= 2.5 kOe (see Fig. 1 (c) and inset of Fig. 2 (a)) and changes sign when the direction of $H_{ext}$ is reversed (ii) Its amplitude monotonously increases with the pump energy density. This phenomenon is due to the modulation of the reflectivity signal by hot-electron pulse and the strain pulse which affects differently the right (σ+) and left (σ-) helicity of light [47]. Since the Kerr rotation can be considered as the phase difference between the reflected σ+ and σ- helicity, the different effect induced in $σ^+$ and $σ^-$ is observed in the $\Delta\Theta_K(t)$. From a theoretical point of view, it can be reproduced within the thin-film multilayer reflectivity model based on the transfer matrix method [47,48]. Such a full theoretical study goes however beyond the scope of the present paper. Interestingly, after the acoustic pulse leaves the Bi-YIG layer, two resonance modes are clearly revealed by the oscillations shown in the $\Delta\Theta_K(t)$ signal with the frequencies of 6.4 and 63.7 GHz, as seen in the Fourier transform spectrum displayed in the inset of Fig. 2(a). As demonstrated hereafter, the first mode is the ferromagnetic resonance mode ($f_{fmr}$) observed via acoustic pulse induced changes of magnetocristalline anisotropy, whereas the second mode ($f_{acous}$) results from the modulation of the MO effect by the propagation of the acoustic pulse in the GGG substrate.

The obtained results suggest that an acoustic strain excitation is at the origin of the observed resonance modes. The existence of such a strain pulse travelling through the sample has been experimentally confirmed by measuring the pump-induced changes in the reflectivity signal $\Delta R(t)/R$ [Fig. 2(b)], which shows oscillations for the time delay higher than 40 ps. Such oscillations were attributed to the so-called Brillouin oscillations, which are due to the interference between the probe beam reflected at the Bi-YIG interfaces and secondary beams reflected by the strain pulse propagating in the GGG substrate. The frequency associated with the Brillouin oscillations is given by $f_B = 2V_{GGG}\sqrt{n^2 - \sin^2\theta}/\lambda$ [49], where λ = 400 nm, $V_{GGG}$ = 6400 m/s [50], n ≈ 2 [45], and θ = 6° are, respectively the probe wavelength, the longitudinal sound velocity in GGG, the refractive index at the probe wavelength and the incidence angle of the probe beam. The calculated value of $f_B$ = 63.9 GHz, which is in good agreement with the frequency characterizing the oscillations observed in $\Delta R(t)/R$ signal [see inset of Fig. 2(b)].

The comparison between the results obtained from $\Delta\Theta_K(t)/\Theta_{Kmax}$ and $\Delta R(t)/R$ measurements allows us to conclude that the mode $f_{acous}$ observed in the TR-MOKE is related to the propagation of the acoustic pulse in the GGG substrate, since it has the same frequency as the Brillouin oscillations. This result clearly demonstrate that an acoustic pulse can induce in a MO medium a structure with a complex refractive index that allows the modulation of the MO effects at a frequency determined by the sound velocity as mentioned by Subkhangulov et al [51]. On the other hand, the frequency of 6.4 GHz associated with the low-frequency mode is in the range of the ferromagnetic resonance (FMR) frequencies in Bi-YIG. In order to confirm the magnetic origin of this mode, we investigated the effect of the external magnetic field on the TR-MOKE. $\Delta\Theta_K(t)$ measured at selected external magnetic field are displayed in Fig. 3(a). The frequency and amplitude of the low-frequency mode are clearly influenced by the external magnetic field. To further highlight the behavior of the modes, the field dependence of the oscillations frequency and amplitude as determined by Fourier analyses are shown in Figs. 3(b) and 3(c). The variation of the oscillations frequency of the low-frequency mode can be described by Kittel formula adapted to the case of our experimental configuration [52]:

$$\omega = \gamma \left(H_{ext} - H_{eff}\right) \quad (1)$$

where $\omega$ is the angular precession frequency, $\gamma$ the gyromagnetic ratio, $H_{ext}$ the external magnetic field, and the effective field $H_{eff}$ is defined as ($4\pi M_s - H_u - H_c$) where $H_u$ and $H_c$ are the uniaxial and cubic anisotropy fields, respectively. The adjustment of Fig. 3(b) with Eq (1) using $\gamma = 2.8$ GHz/kG yields $H_{eff} = 1.22$ kG. This behavior of the low-frequency mode clearly indicates that is associated with FMR. We also note that we have found that the initial precession amplitude for the low-frequency mode has maximum near the saturating field $H_{sat} = 2.5$ kG (see Fig 1(c) and Fig (3c)), which is also in a qualitative agreement with the typical behavior obtained for the FMR mode when $H_{ext}$ is applied along a hard magnetization axis as in our experimental configuration [53,54]. On the other hand, the frequency of $f_{acous}$ is independent on the magnetic field strength (Fig. 3(c)). As discussed above, this is the expected behavior for the acoustically-induced modulation of the MO effects. Furthermore, we show that the initial oscillations amplitude of $f_{acous}$ as a function of field has the same behavior as the MO response of our sample. Such dependence can be related to the characteristic behavior of the non-magnetic contribution observed in $\Delta\Theta_K(t)$ for t between 0 and 40 ps which modulate the TR-MOKE signal.

Let's us now focus on the mechanism behind the excitation of the ferromagnetic resonance in our sample. It results from an ultrafast non-thermal modification of the magnetocrystalline anisotropy induced by the acoustic strain pulses via the inverse magnetostriction effect, as initially shown by Scherbakov et al in GaMnAs [18]. The excitation of the FMR mode for a magnetization already aligned along $H_{ext}$, i.e higher than $H_{sat}$ = 2.5 kG (see Fig. 1(c) and Fig. 4), substantiate this interpretation. Indeed, a decrease of the magnetocrystalline parameters induced by heating effects does not allow exciting the FMR mode when the magnetization is aligned along $H_{ext}$ [25]. Let us also mention that the non-thermal mechanisms based on the Cotton-Mouton effect [55] and photo-induced magnetic anisotropy [12,32] usually used to induce in magnetic garnet a magnetization precession with a linearly polarized light can be excluded in our case. This is due to the very weak transmitted light (less than 0.1%) from the thick Pt/Cu bilayer to the Bi-YIG layer. Moreover, we have investigated using the same configuration the ultrafast magnetization dynamics induced by direct light excitation (not shown). No magnetization precession has been observed. This results further highlights the importance of the approach based on acoustic strain pulse for generating spin wave via the inverse magnetostriction effect.

To further investigate the two resonance modes, we performed TR-MOKE measurements as a function of the laser energy density $E_{pump}$. Figure 4(a) shows the TR-MOKE signals measured at selected $E_{pump}$ for $H_{ext}$= 3.3 kG. The pump energy density dependence of the oscillations frequency and amplitude extracted using Fourier analysis are presented in Figs. 4(b) and 4(c). The frequency of both modes is independent of the $E_{pump}$. The behavior of $f_{fmr}$ is similar to the one obtained by non-thermal effects induced magnetization precession [32,34]. This is in agreement with our interpretation based on strain-induced changes of magnetic anisotropy via the inverse magnetostriction effect. Indeed, in the case of thermally induced spin precession a dependence of the frequency on $E_{pump}$ is usually observed [56,57]. On the other hand, the behavior of $f_{acous}$ as a function of $E_{pump}$ is also in agreement with the prediction that the frequency of acoustically-induced modulation of the MO effects is mainly defined by the speed of sound. Moreover, our experiments show that the oscillations amplitude of the two resonance modes increases linearly with the laser energy density within the probed range. This means that the amplitude of the spin precession is proportional to the amplitude of the strain pulse. Therefore, using an engineered structure that allows injecting a higher amplitude strain pulse into Bi-YIG can be used for further improving the magnetization precession amplitude or inducing a magnetization switching in magnetic garnet.

## VI. CONCLUSION:

We have studied the laser-induced ultrafast magnetization in a dielectric film of bismuth-substituted yttrium iron garnet buried below a thick Pt/Cu bilayer. It is found that exciting the sample from Pt surface launches coherent stain pulses that propagate into the garnet film. We demonstrate that this acoustic pulse modifies the magnetocrystalline anisotropy via the inverse magnetostriction effect. This triggers a coherent magnetization precession at the frequency of the ferromagnetic resonance. Importantly, we can control the amplitude of the spin precession by tuning the amplitude of the acoustic strain pulse. Our results highlight the suitability of acoustic strain pulse for generating spin wave in dielectric materials.

**ACKNOWLEDGMENTS**

The authors acknowledge M. Bargheer for interesting discussions. This work was supported by the ANR-NSF Project,ANR-13-IS04-0008-01, COMAG, ANR- 15-CE24-0009 UMAMI and by the ANR-Labcom Project LSTNM, by the Institut Carnot ICEEL for the project « Optic-switch » and Matelas and by the French PIA project 'Lorraine Université d'Excellence', reference ANR-15-IDEX-04-LUE. Experiments were performed using equipment from the TUBE. Davm funded by FEDER (EU), ANR, Région Grand Est and Metropole Grand Nancy.

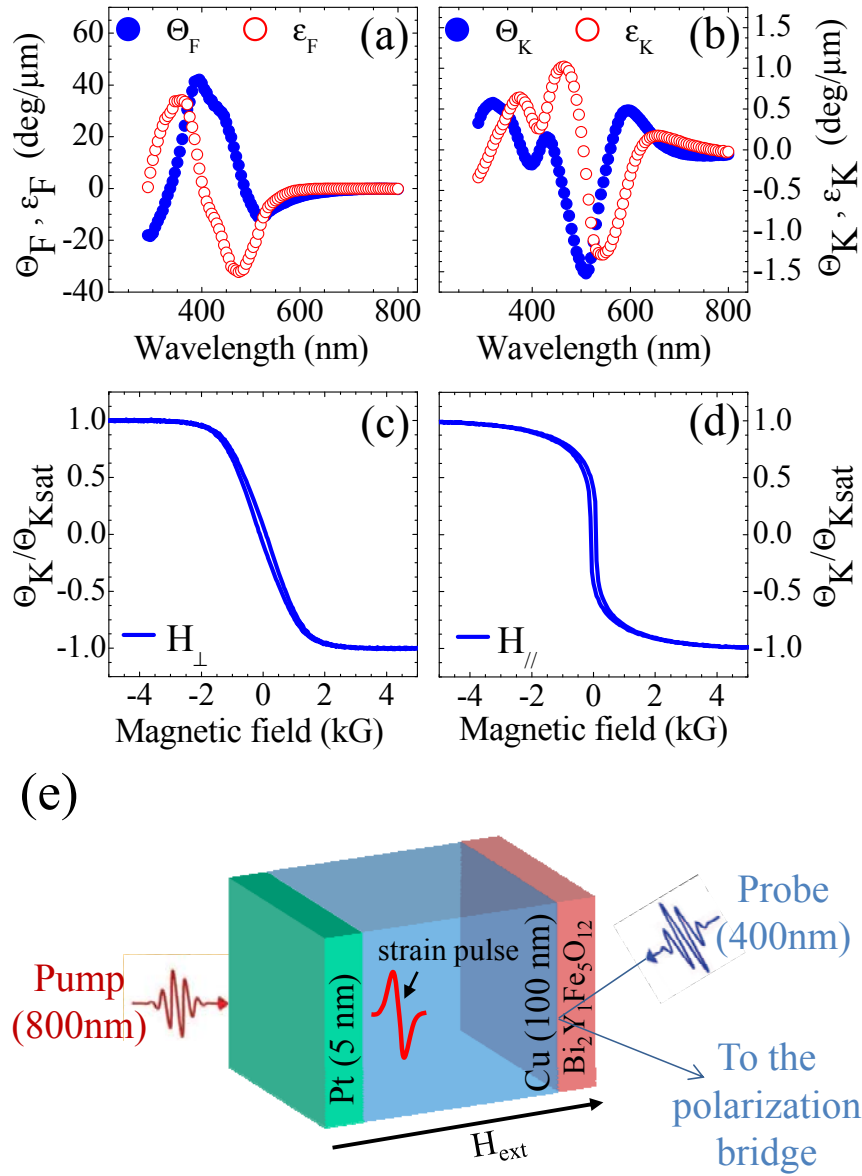

Figure 1: Static room temperature magneto-optical and magnetic properties of $Bi_2Y_1Fe_3O_{15}$ thin film and the pump-probe experimental configuration. (a, b) Magneto-optical Faraday (a) and Kerr (b) polar spectra measured over a broad range of wavelength. The filled and open symbols represent, respectively, the rotation ($\Theta_F$, $\Theta_K$) and ellipticity ($\varepsilon_F$, $\varepsilon_K$). (c,d) Normalized magneto-optical hysteresis loops measured in polar (c) and longitudinal configuration. (e) Sketch of the time resolved experimental configuration that allows studying the ultrafast magnetization dynamics induced by acoustic pulse.

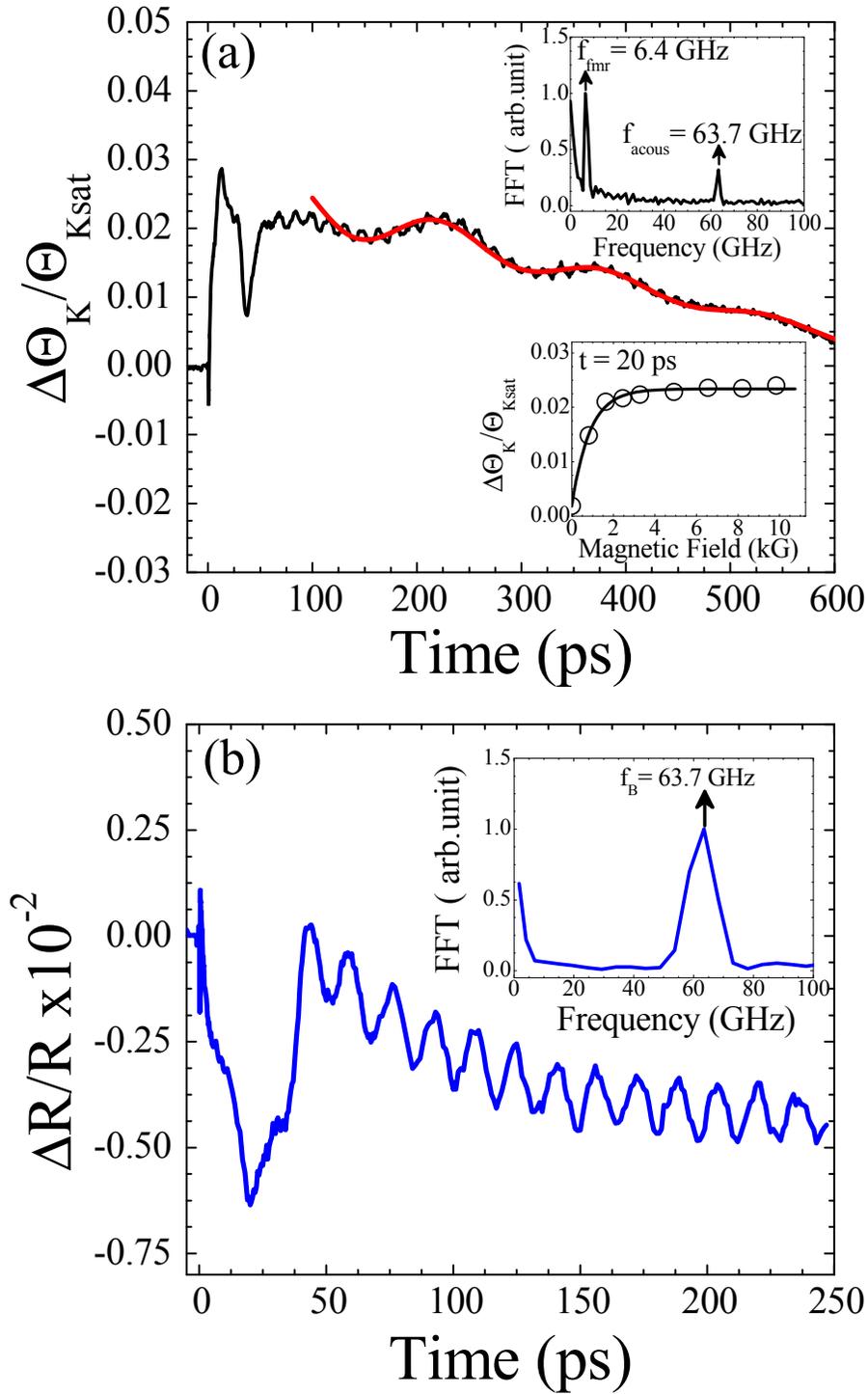

Figure 2: Dynamics of spin and reflectivity in the $Bi_2Y_1Fe_5O_{12}$/GGG(100) buried below a thick Pt/Cu bilayers. (a, b) $\Delta\Theta_K/\Theta_{Ksat}$ and $\Delta R/R$ induced by a laser energy density of 11.3 mJ cm$^{-2}$ for $H_{ext}$= 3.3 kG. Inset (a): the Fourier transform spectrum of the $\Delta\Theta_K/\Theta_{Ksat}$ data for the time delay t ≥ 50 ps and the $\Delta\Theta_K/\Theta_{Ksat}$ measured at the time delay t = 20 ps as a function of $H_{ext}$. Inset (b): the Fourier transform spectrum of the $\Delta R/R$ data for the time delay t ≥ 50 ps. The solid lines in (a) are guides to the eyes.

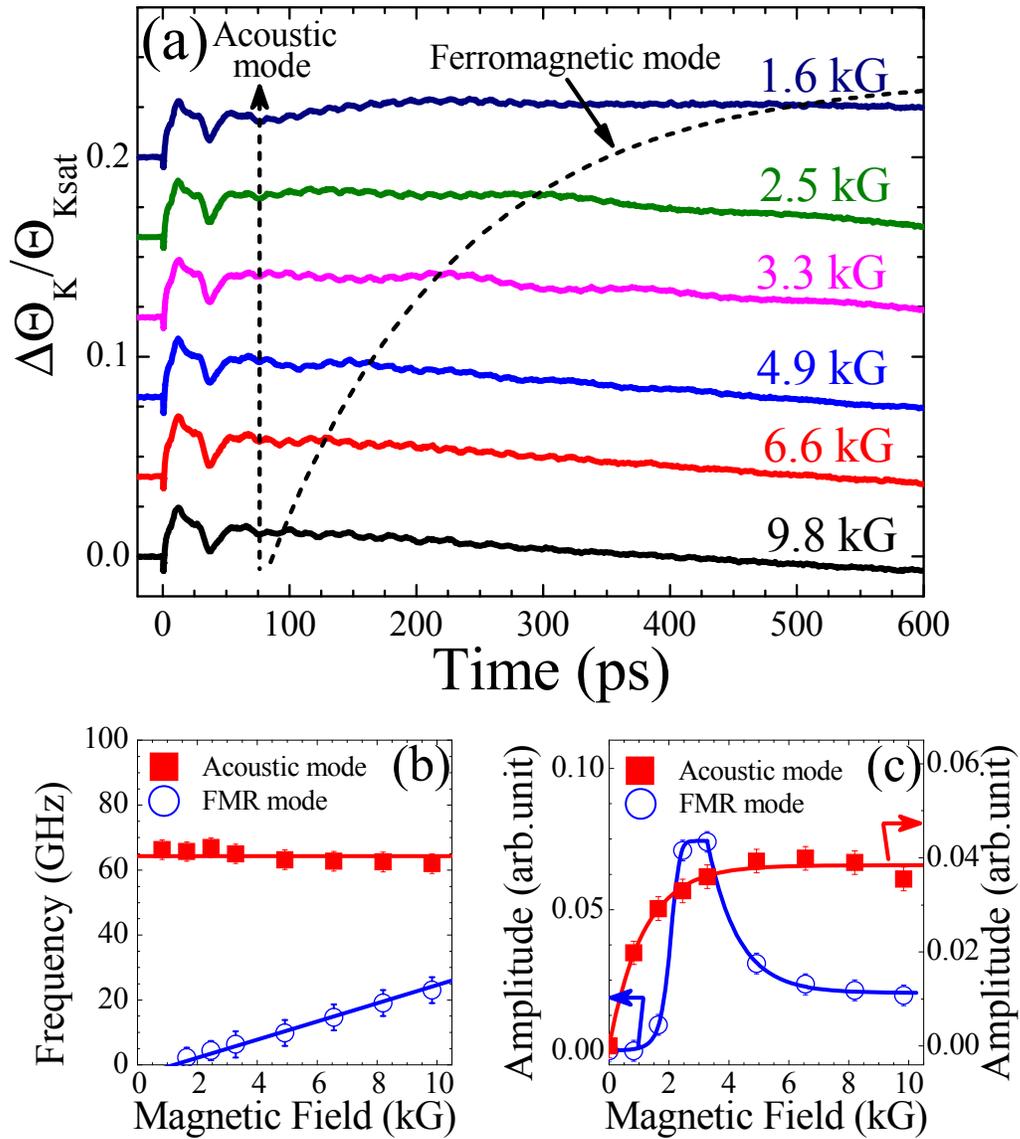

Figure 3: Magnetic field dependence of the spin dynamics. (a) $\Delta\Theta_K/\Theta_{Ksat}$ as a function of the magnetic field. (b, c) Field dependence of the precession frequencies (b) and amplitudes (c) associated with acoustic and FMR resonance modes. All measurements are obtained for a pump energy density of 11.3 mJ cm$^{-2}$. The dashed lines in (a) and solid lines in (c) are guides to the eyes. In (b) the solid line describing the FMR mode is a fit with the Kittel formula, whereas the solid line for the acoustic mode is a guide to the eyes.

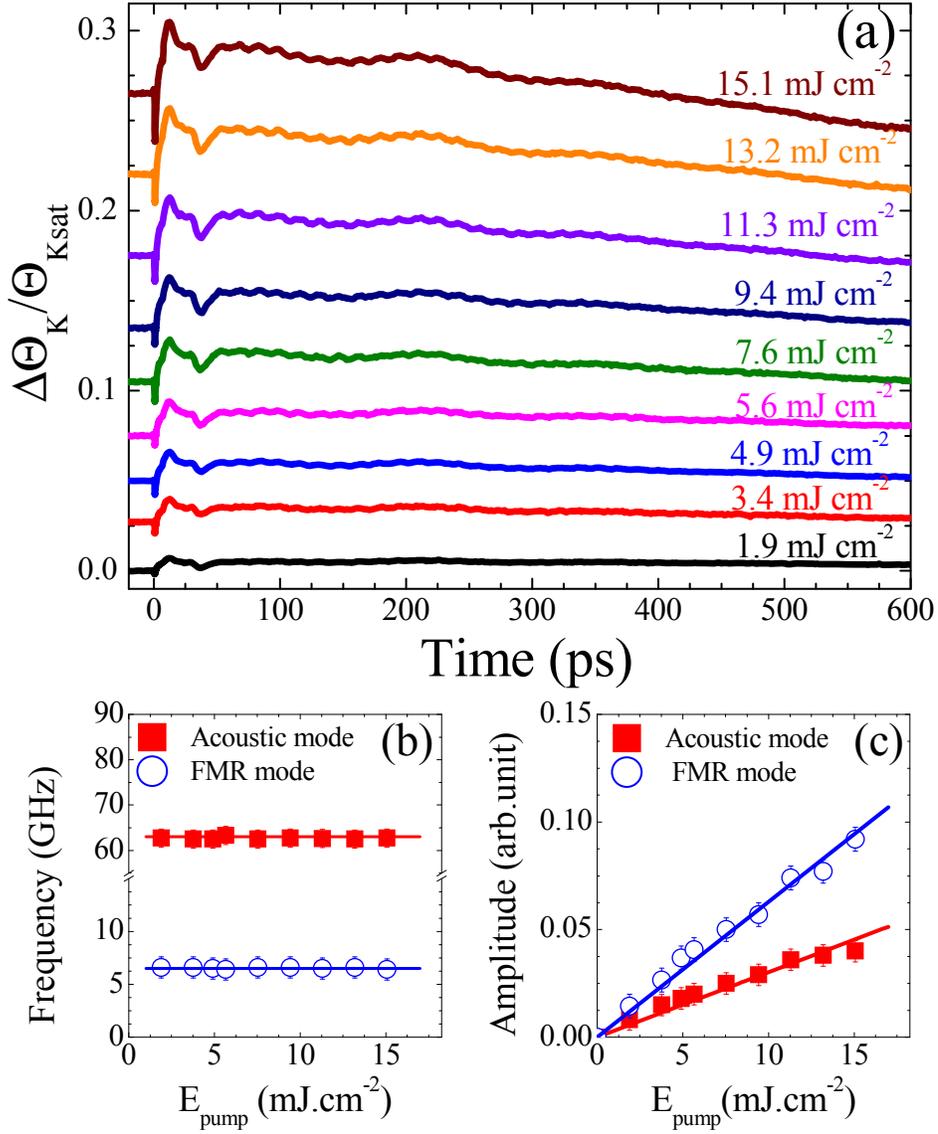

Figure 4: Laser energy density dependence of the spin dynamics. (a) $\Delta\Theta_K/\Theta_{Ksat}$ as a function of the laser energy density. (b, c) Variation of the precession frequencies (a) and amplitudes (b) associated with acoustic and FMR resonance modes as a function of the laser energy density. All measurements are obtained for $H_{ext}$ = 3.3 kG. The solid lines in (b) and (c) are guides to the eyes.